# Proposal of ROS-compliant FPGA Component for Low-Power Robotic Systems

## －case study on image processing application－

Kazushi Yamashina, Takeshi Ohkawa, Kanemitsu Ootsu and Takashi Yokota
Graduate School of Engineering, Utsunomiya University
Yoto 7-1-2, Utsunomiya, Tochigi, Japan, 321-8585
kazushi@virgo.is.utsunomiya-u.ac.jp +81-28-689-6270

***Abstract* — *In recent years, robots are required to be autonomous and their robotic software are sophisticated. Robots have a problem of insufficient performance, since it cannot equip with a high-performance microprocessor due to battery-power operation. On the other hand, FPGA devices can accelerate specific functions in a robot system without increasing power consumption by implementing customized circuits. But it is difficult to introduce FPGA devices into a robot due to large development cost of an FPGA circuit compared to software. Therefore, in this study, we propose an FPGA component technology for an easy integration of an FPGA into robots, which is compliant with ROS (Robot Operating System). As a case study, we designed ROS-compliant FPGA component of image labeling using Xilinx Zynq platform. The developed ROS-component FPGA component performs 1.7 times faster compared to the ordinary ROS software component.***

## I. Introduction

In recent years, the design techniques for building robots are actively studied. In addition, robots (e.g., for disaster relief, unmanned drone, and so on) are required to be autonomous and their robotic software are sophisticated. The robots cannot equip with high-performance processor since they are expected to be in battery operation. On the other hand, FPGA (Field Programmable Gate Array) devices contribute to speed up in fields such like network packet routing and image processing. FPGAs can employ parallel operations for specific functions while software always runs sequentially. But a development of a system using an FPGA is more difficult than that of software since there is a need for implementing them with Hardware Description Language (HDL). High-level synthesis of hardware is also emerging in recent years to implement circuit with C language, however, it is still difficult for ordinary software engineers to handle them.

Robotics depends on various expertise, so they become to be hard to develop [1]. Therefore, the development of systems using FPGA devices needs to reduce cost for easy integration of FPGAs into robots. Component-oriented development is a well-known method for reduction of costs in the development of robotic software. ROS (Robot Operating System) has been proposed as a software platform of component-oriented development of robots [2]. ROS provides a framework of communication layer and a build system for robotic software.

In this study, we propose a hardware component using an FPGA for easy integration of an FPGA into robots, which can be complied and used in ROS system. The ROS-compliant FPGA component contributes to performance improvement of robots using FPGA devices without lowering productivity of robot developers.

This paper presents a practical design of ROS-compliant FPGA component that performs labeling process of images as an example. The component is implemented on a programmable SoC. In addition, the performance of ROS-compliant FPGA component was evaluated by comparing with software.

## II. "ROS" Framework for Robot

### A. ROS Overview

ROS (Robot Operating System) is released by OSRF (Open Source Robotics Foundation) as an open source project [2]. It is a software platform which provides a framework of communication layer and a build system for robotic software. ROS runs mainly on Linux. Motivation of ROS is to support for *reuse* software and to build robotic systems with software component in robotic research and the development. Figure 1 shows the increase of the number of ROS software packages. The number of packages increases rapidly since released in 2007. In addition, Table 1 shows a number of software packages and citations for several software platforms for robots. It is clear that ROS is becoming a kind of mainstream.

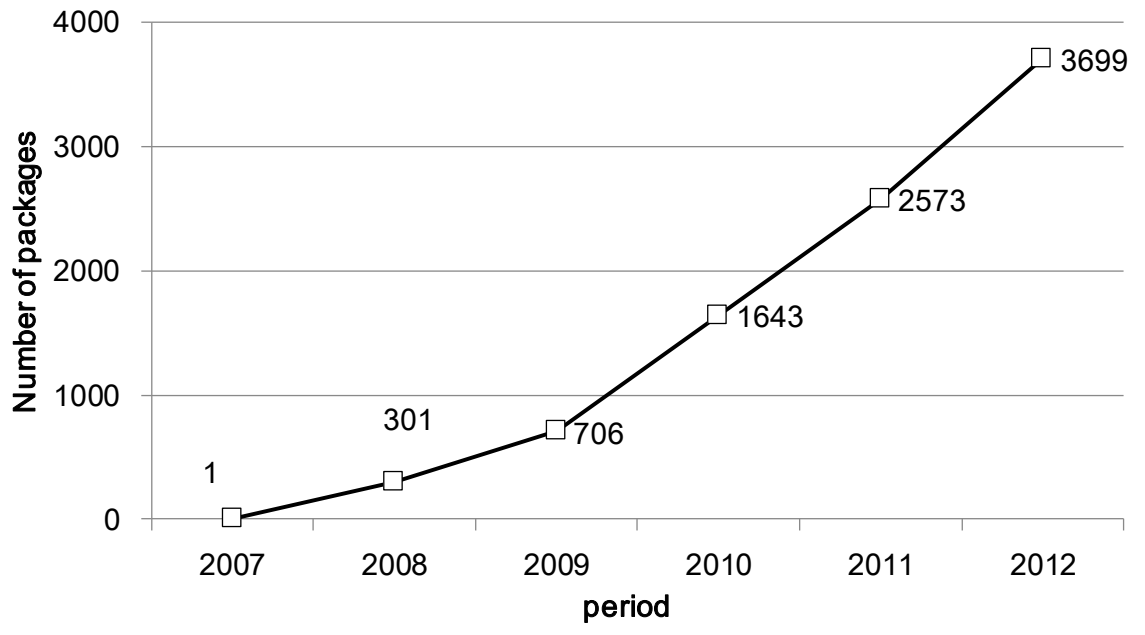

**Figure 1 Increase of ROS software packages**

**Table 1 Number of registered packages and citations for each software platform for robots**

| *Platform Name* | *Packages* | *Citations** |
|---|---|---|
| ROS | 3699 | 4610 |
| RT-middleware [3] | 321 | 1090 |
| OROCOS [4] | Unknown | 1540 |

*google scholar May 5th, 2015

### B. ROS communication model

In ROS development, a robot system is designed using a set of component called "*node*" and its communication channel called "*topic*". A robot developer can make a robot system by collecting components from a number of distributed package and connecting them. The developer can also make a *node* when a new function is needed.

The communication model of ROS is based on Publish-Subscribe messaging (Figure 2). Publish-Subscribe messaging is an asynchronous messaging model that ROS Nodes communicate over a topic to each other. The biggest advantage of Publish-Subscribe messaging is a dynamic network configuration since the ROS nodes are bound loosely. Therefore, it is able to add a new ROS node easily.

There are two roles in ROS nodes: *publisher* and *subscriber*. *Topic* is a classification of message as well as a name of the communication path. A *publisher* node publishes a message to a topic. The topic holds a sequence of messages. And any *subscriber* node in the system, which has subscribed to the topic in advance, can receive the message.

*Publisher* nodes do not know about *subscriber* nodes. In other words, ROS nodes do not have information of communication partner and are bound loosely. Therefore, ROS is able to be dynamic network configuration.

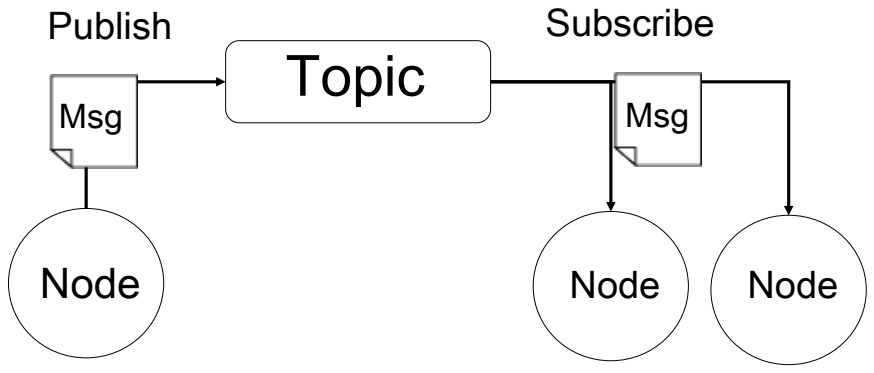

**Figure 2 Publish - Subscribe messaging**

## III. ROS-COMPLIANT FPGA COMPONENT

This section describes the requirements for ROS-compliant FPGA component and its target hardware platform.

### A. Requirements of ROS-compliant FPGA component

First, we define "*ROS-compliant"* as follows: An FPGA component is ROS-compliant when the component conforms to publish/subscribe messaging rule so that it can communicate with any other ROS nodes. There are two requirements for ROS-compliant FPGA component.

- The functionality of the ROS-compliant FPGA component is equivalent to that implemented in software.
- The message type and data format used at the input and output of the ROS-compliant FPGA component is equivalent to that implemented in software

An integration of an FPGA into a robotic system needs equivalent functionality to replace a software ROS component with a ROS-compliant FPGA component. Therefore, each ROS message type and data format used in ROS-compliant FPGA component must be same as the software ROS component. ROS-compliant FPGA component aims to improve its processing performance while satisfying their requirements.

### B. Structure of ROS-compliant FPGA component

Figure 3 shows the structure of the proposed ROS-compliant FPGA component model. Based on the requirements in the previous section, the component must implement following four functions:

- The encapsulation of FPGA circuits,
- Interface between ROS software and FPGA circuits,
- *Subscribe* interface from a topic, and
- *Publish* interface to a topic.

The FPGA part performs any accelerated processing. There are two roles of software in the component. First, *an interface process for input* subscribes to a topic to receive input data. It is responsible to format the data suitable for the FPGA processing and sends the formatted data to the FPGA. Second, *an interface process for output* receives processing results from the FPGA. It is responsible to format the data suitable for ROS system again, and publishes them to a topic. If other nodes need the data on the topic, the nodes should have subscribed to the topic. Such a structure realizes to make a robot system in which software and hardware cooperate.

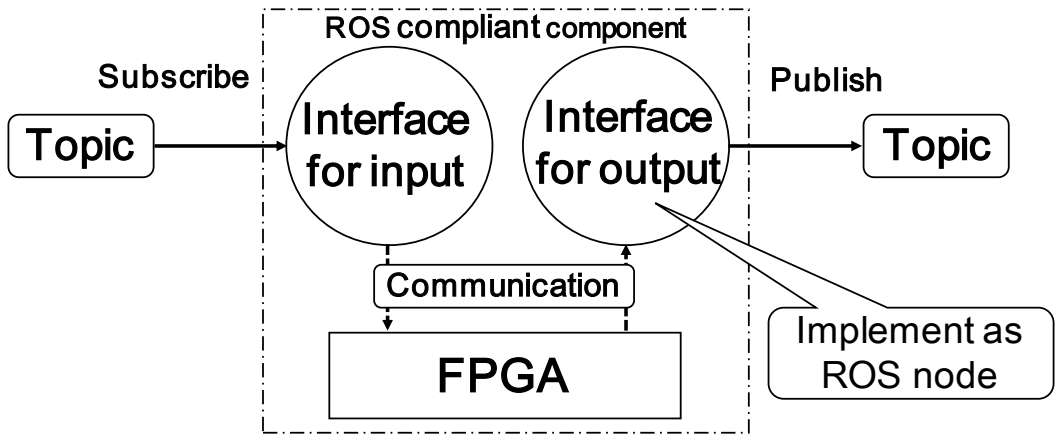

**Figure 3 ROS-compliant FPGA component model**

### C. Implementation on a Programmable SoC

The difference of ROS-compliant FPGA component from a ROS node written in pure software is that processing contains hardware processing of an FPGA or not. Integration of ROS-compliant FPGA component into a ROS system only needs to connect to ROS nodes by Publish/Subscribe messaging in ordinary ROS development style. ROS-compliant FPGA component provides easy integration of an FPGA by wrapping it with software.

We have selected "Programmable SoC" as a target hardware platform for low-power robots. Figure 4 shows assignment of processing on a programmable SoC. SoC (System on Chip) is an LSI that integrates all of the necessary functions to build a system in a single chip. Generally, a hardwired processing logic on an FPGA can achieve better performance at low power consumption compared to software running on a general purpose processor. T. Suzuki et al. exhibited power saving and performance improvement of processing of robotic application using an FPGA [5]. In this study, we use ARM processor and FPGA in programmable SoC. Linux OS and ROS nodes run on the ARM processor, and processing for robot runs on FPGA. Processing which is suitable for software is assigned to software on ARM, and processing which is suitable for hardware is assigned to hardware. The aim of choosing “Programmable SoC” as a target platform is to improve performance by minimizing burden of hardware development while keeping productivity.

## IV. CASE STUDY : LABELING COMPONENT

This section describes implementation of image labeling based on ROS-compliant FPGA component model.

### A. *Labeling overvew*

Processing of image labeling that it puts label number each group of white pixels in binary image. Labeling is used to measure area, angle and length of target in many robotic systems. Figure 5 shows an example of labeling result. To produce a correct result, the labeling need two steps, because raster-scan may label one region as a split several regions. The second step may fix the problem easily. Anyway, this paper describes the first step of labeling, which tends to be time consuming task for software.

### B. *Hardware implementation overview*

Figure 7 shows the block diagram of the hardware. Processing target is full HD image (1920×1080 pixels, about 2M bytes). In this case, labeling operation is applied per line basis, since BRAM of an FPGA is too small to store the entire HD image. In order to exploit maximum efficiency of hardware processing on an FPGA, we designed labeling hardware to be able to label a pixel in a clock.

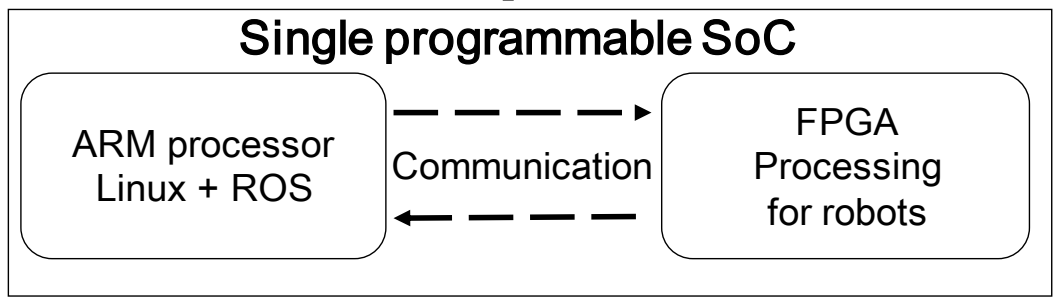

**Figure 4 Assignment of tasks on a programmable SoC**

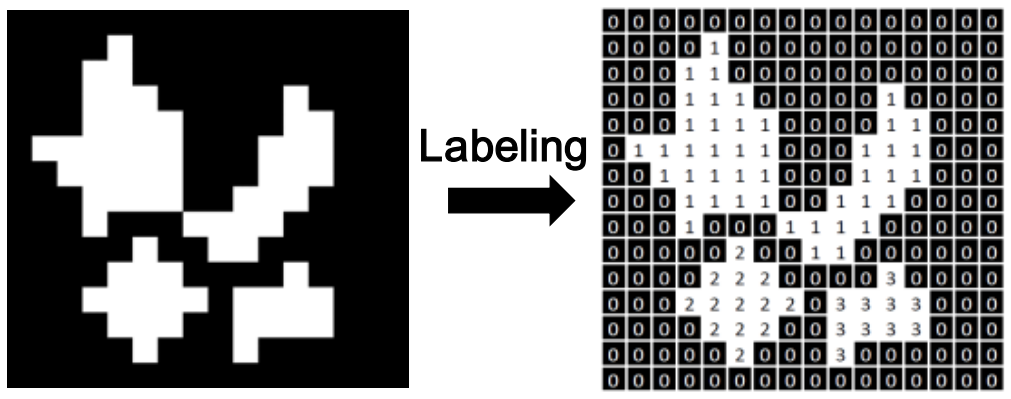

**Figure 5 Labeling example**

#### 1) *Communication between FPGA and ARM processor*

In this study, Xillinux [6] is used to communicate between FPGA logic and ARM processor. Xillinux is a platform for Zynq that is released by Xillybus Ltd. Linux (Ubuntu) OS runs on the ARM processor. Xillinux can access to FPGA logic through a specific device file.

Figure 6 shows the communication mechanism of the labelling hardware and the ARM processor. Software can access to FIFOs through device file and read/write data from/to it. The FPGA reads/writes data from/to FIFO by control of read/write enable port at any time.

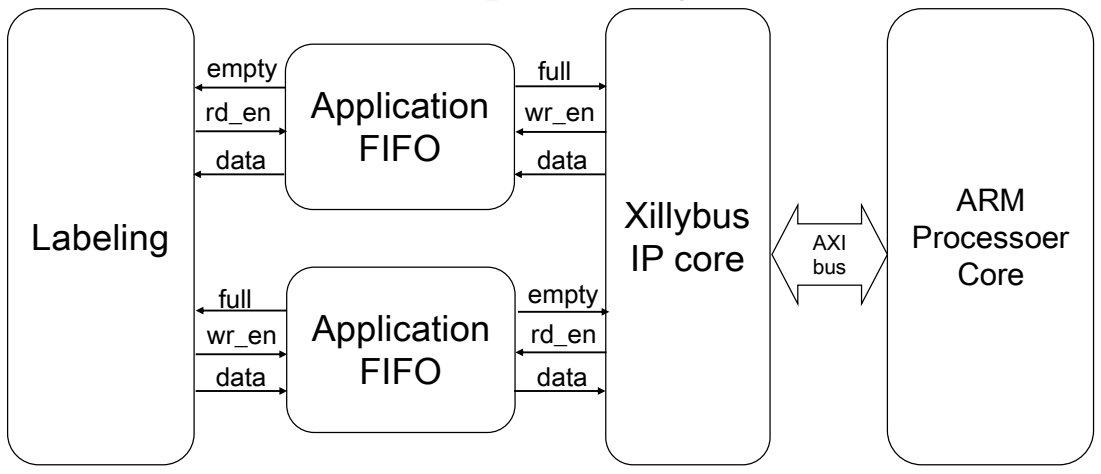

**Figure 6 Communication mechanism**

#### 2) *The role of each module*

Table 2 shows the roles of each module in hardware. There are five modules. First, the labeling hardware stores a line of input image to *memory_img*. Immediately after that, *input_controller* inputs pixel data to *label_generator* and l*abel_generator* executes the labeling pixel by pixel. Labeling algorithm needs a pixel data and label numbers of the previous line and previous pixel, so two memories *label_data0* and *label_data1* are prepared. In other words, *label_generator* writes result of current line to *label_data1* while reading from *label_data0*. In the next line, *label_generator* read from *label_data1* and write to *label_data0*.

Detail of “label_generator” is explained in the next section.

**Table 2 Function of each module**

| *Module* | *Function* |
|---|---|
| memory_img | Save a line of input image |
| label_generator | Labeling a pixels |
| label_data0 | Save a line of result (label number) |
| label_data1 | Save a line of result (label number) |
| FIFO buffer | Buffer for input and output |

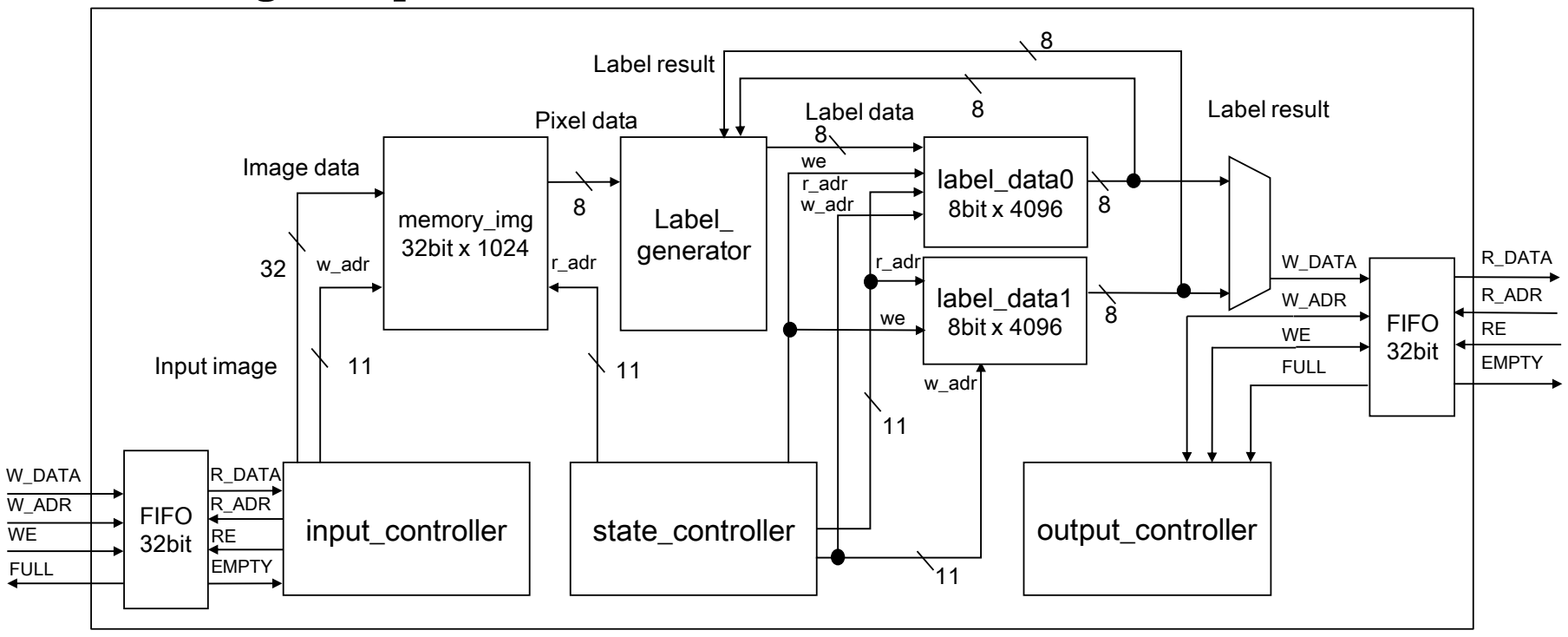

**Figure 7 Overall view of hardware**

*3) Detail of labeling curcuit (label_generator)*

We have designed labeling hardware to be able to label a pixel in a clock. Figure 8 shows a circuit diagram of label_generator. There are four 8-bit inputs. One input named "New Pixel" is for pixel data of input image, other inputs are for previous label numbers. In addition, single output named "Output Label" is 8-bit for labeled number.

Whenever "New Pixel" is black ("0"), the output is "0". When it is white, if "*Reference Labels*" are all "0", the circuit outputs an incremented value of previous label number, which is stored in the "*Current Label*" register. On the other hand, if there are any non-zero number in "*Reference Labels*", minimum number of them is output as a result.

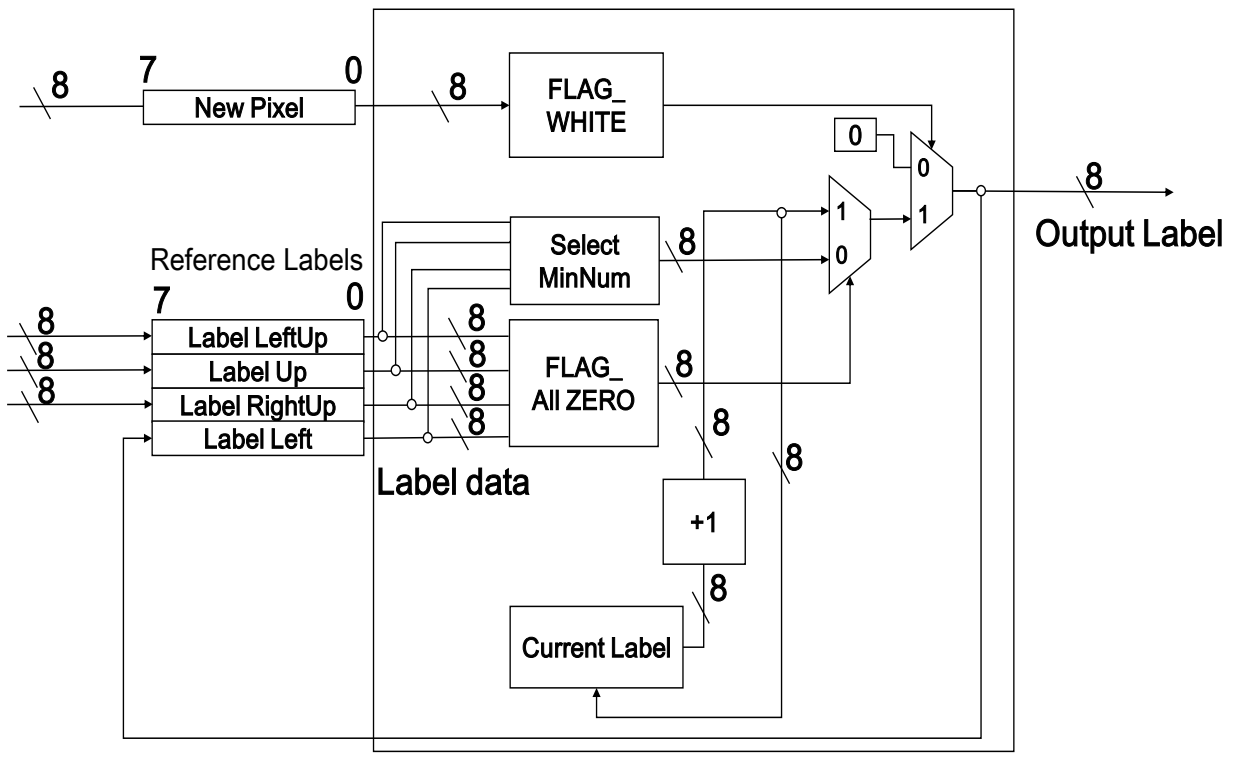

**Figure 8 Detailed block diagram of *label_generator***

*C. ROS node implementation overview*

Figure 9 shows overall view of ROS system which includes ROS-compliant FPGA component. There are four ROS nodes in the system.

- *input_image*: input an image to *data_input (topic)*
- *write2fpga*: receive data from subscribed *data_input* and send it to an FPGA
- *read4fpga*: receive data from an FPGA and publish it to *data_output (topic)*
- *display_result*: receive data form subscribed *data_output* and display it on a console.

*Input_image* reads input image from a bitmap file and publishes it to topic *data_input*. In *write2fpga,* message received from the subscribed topic *data_input* is sent to FPGA as image data. *Write2fpga* for providing data to the FPGA accesses FIFO through device file and writes data. After labeling on the FPGA, *read4fpga* reads data from the FPGA, similarly. Then the label number data read from the FPGA are published to *data_output* as message. If other node needs labeling results, any node in the system can subscribe to the topic and receive the result from it.

The format of ROS message can be defined at message file like Figure 10 by a developer. A developer can use pre-defined message prepared by ROS, too. In this study, we have defined four fields in the message used in the system.

- *frame_ID*: frame number (32bit integer)
- *width* : width of image (16bit integer)
- *height* : height of image (16bit integer)
- pixels : for pixels data of image (32bit integer array)

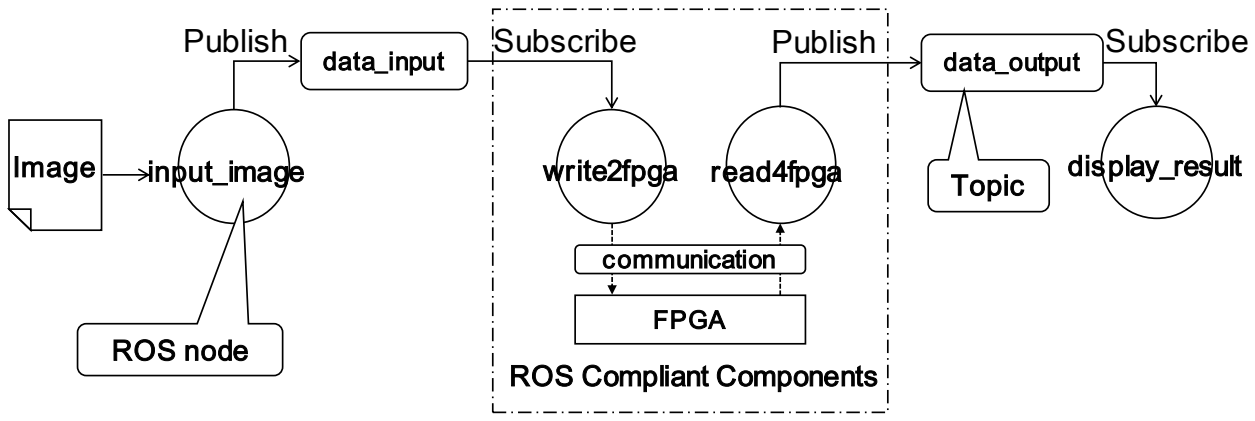

**Figure 9 Overall view of ROS system with ROS Compliant FPGA Component**

```
int32 frame_ID
int16 width
int16 height
int32[ ] pixels
```

**Figure 10 Message file used to define message format**

## V. Performance Evaluation

This section describes performance evaluation of ROS-compliant FPGA component. The evaluation was conducted in three different conditions.

(1) ROS-compliant FPGA component (ARM + FPGA)
(2) Software only (SW only : ARM)
(3) PC (SW only : PC)

The environment used for (1) and (2) was ZC7Z020 (Xilinx Ltd) on Zedboard. ZC7Z020 is a programmable SoC equipped with ARM Cortex-A9 processor (666MHz) and Artix-7 FPGA on a chip. OS is Ubuntu 12.04 LTS (Xillinux-1.2-eval). In addition, the operating frequency of labeling hardware was set to 100 MHz in (1). The environment of (3) is ordinary PC equipped with Intel Core i7 870 (2.93GHz). OS is Ubuntu 12.10. In addition, Table 3 shows hardware resource utilization of the FPGA.

Figure 11 shows the average of measured processing time in labeling. Resolution of input image is 1920x1080, and the measurement was done using gettimeofday() standard C library function in the software and repeated 10 times. In the environment of (1), the average was 32 ms per frame (*including communication time between the ARM processor and FPGA*), the min/max were 28 ms / 35 ms, respectively. The processing time of ROS-compliant FPGA component was 26 times faster than that of software with the ARM processor, and the ROS-compliant FPGA component performs even 2.3 times faster than that of PC.

Ideally, the number of clocks is 1,920 clocks for the processing per line. In the current implementation, the period is 2,400 clocks because 5 clocks is elapsed to process 4 pixels.

**Table 3 Hardware resource utilization (Zynq-7020)**

| *RESOURCE* | *UTILIZATION* |
|---|---|
| Slice Registers | 4,123/106,400 (3%) |
| Slice LUTs | 4,114/53,200 (7%) |
| RAM B36E1 | 3/140 (2%) |
| RAM B18E1 | 11/280 (3%) |

Therefore, processing per frame requires 2.6M clocks for 1,080 lines, and the processing time per frame in the FPGA is 26ms (the clock period is 10 ns). This means overhead of the processing time (32 ms) in the experimental system is 6 ms. This is due to communication time between the ARM processor and the FPGA. The ROS-compliant FPGA component performs faster than processing with only software even with care of the communication delay between the software on ARM processor and the FPGA.

Figure 12 shows the total latency from the data input to the data output in ROS-compliant FPGA component and the ROS component with pure software. This latency corresponds to the performance in real situation of robot system. Legends of represents is as follows.

1: Communication of ROS nodes (Publish/Subscribe)
2: From after subscribe to before labeling
3: Labeling
4: From after labeling to before publish
5: ROS node's communication

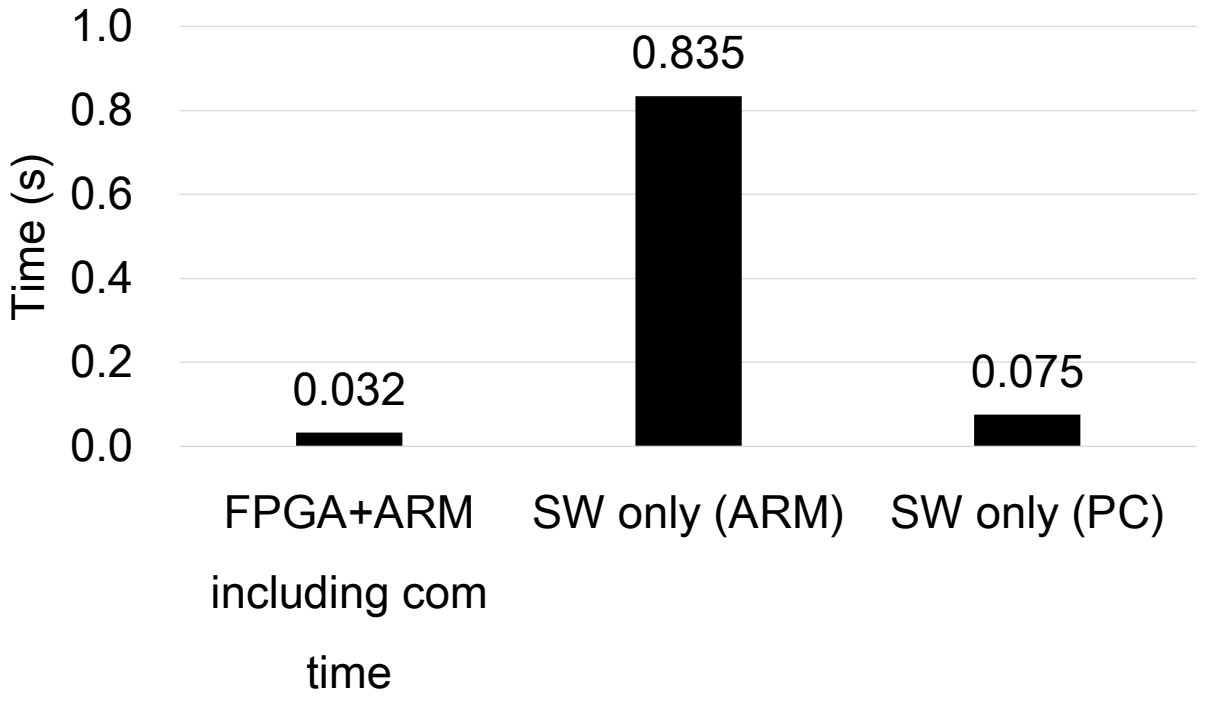

**Figure 11 Measured processing time of labeling**

In the environment (1), the total latency was 1.99 sec. This is about 1.7 times shorter than that of pure software with the ARM processor. In (1) and (2), communication of ROS nodes occupies a lot of the total latency. In addition, the latency of (3) is much shorter than (1) and (2). Constituent ratio of labeling is about the same in (2) and (3). The time of communication and computation is proportional to processor performance. In (1), the time of computation is reduced drastically, so the time of communication is relatively longer. Operating frequency of ARM processor is 666 MHz in (1) and (2), on the other hand, Intel Core i7 870 is 2.93GHz.

Regarding power consumption, the supply power of labeling hardware on the ROS-compliant FPGA component was estimated with XPower Analyzer by Xilinx that is attached to ISE Design Suite. The total power estimated for our ROS-compliant FPGA component is *0.33W*, the dynamic power is 0.20W, and the static power is 0.13W. Generally, power consumption of high performance processor is about 100W (for example, Intel Core i7 and so on). The power consumption of the proposed ROS-compliant FPGA component is much lower than that of the high-performance processors. Table 4 shows the power consumption of several wheel-based vehicle robots by battery operation. They ranges from 22W to 400W. The estimated power of the labeling component is much lower compared to them. Therefore, ROS-compliant FPGA component can contribute to the performance improvement of robots while keeping low power.

**Table 4 Power consumption of robots**

| ROBOT | BATTELY POWER |
|---|---|
| Kobuki [7] | 22W |
| iRobot Roomba [8] | 30W |
| Husky A200 [9] | 400W |

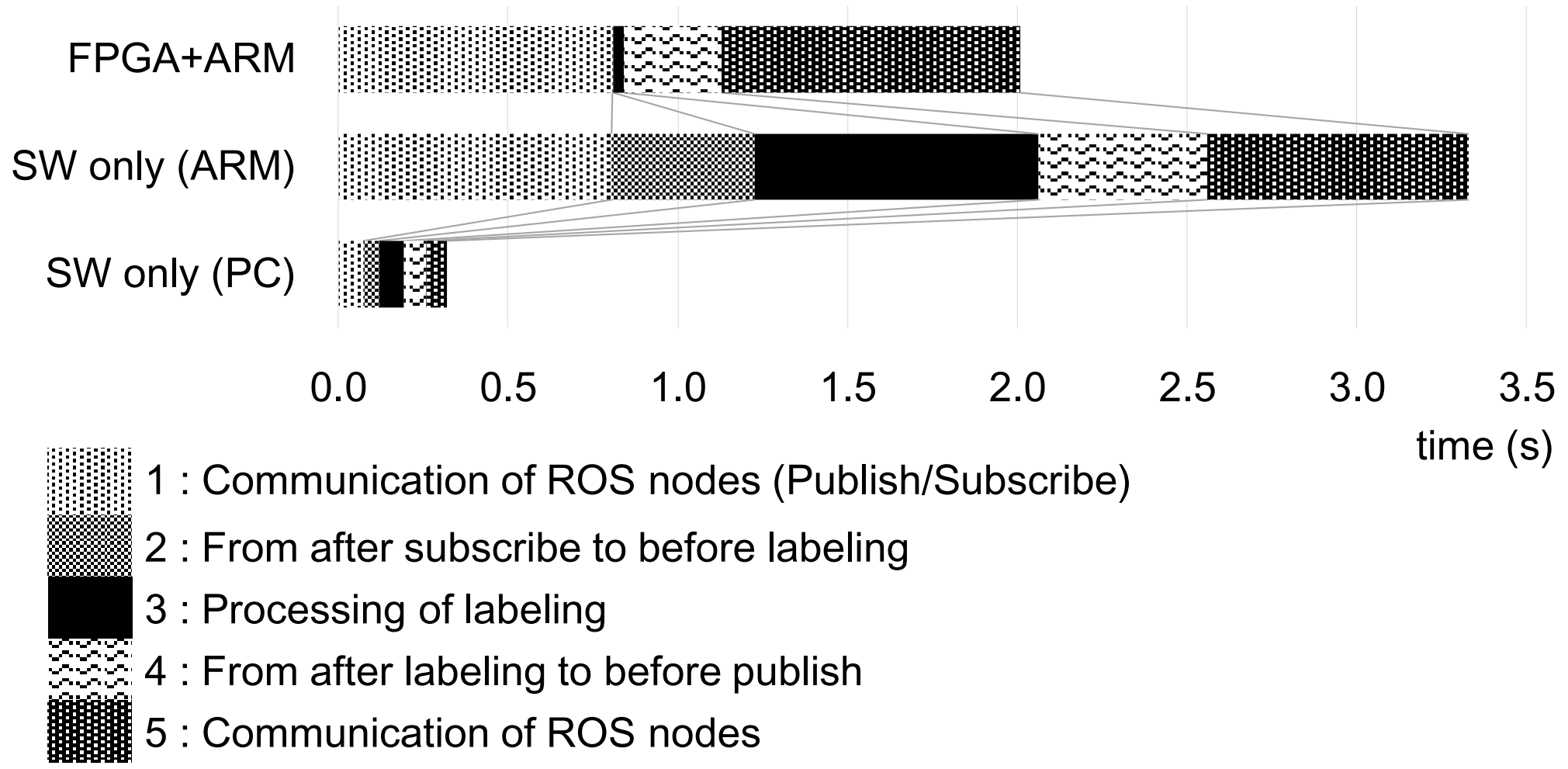

**Figure 12 Total latency of the ROS compliant FPGA component**

## VI. Related works

There are not few papers which report the application of an FPGA onto a robot. This section describes some examples which FPGA is used for robots.

As a digital logic interacting with real-world interface, FPGA is used for robot manipulator with advanced control logic. This is because most of nonlinear controllers need real time mobility operation which is hardly realized by general purpose microprocessors [12] [13].

In previous research [14], autonomous fuzzy behavior control, and sensor-based behaviors are implemented on an FPGA for robot car. These are needed for the human-like driving skills by an autonomous car-like mobile robot.

Another view point of the application of an FPGA onto robotic systems is the design environment of control systems using an FPGA. The development of a system using an FPGA is more difficult than that of software since there is a need for implementing them with Hardware Description Language (HDL). It is still difficult for ordinary software engineers to handle them, so how to reduce the development cost of a system using FPGA is very important. In previous research [16], it is mentioned that the development of a system on an FPGA using traditional programming language: C, C++ and MATLAB improve productivity for robotic developer. Recently, our research group developed an inverted pendulum system using the high level synthesis tool which generates HDL code from pure Java code. The computation time of the control logic of the system is greatly reduced by using FPGA designed with the Java code, without writing any HDL code [17].

FPGA is very effective for robots to reduce computation but connection between software and FPGA is not easy. This paper proposed ROS-compliant FPGA component as a technology which build easy a bridge between software and FPGA. Robotic developer can choose any language to implement robotic software and hardware by ROS-compliant FPGA componentizing processing for robots. Once implementing with the ROS-compliant FPGA component, it can achieve performance improvement of robotic system.

## VII. Conclusion

This paper describes ROS-compliant FPGA componentizing of image processing hardware on a programmable SoC. The proposed ROS-compliant FPGA component technology using FPGA devices is aimed to contributes to the easy integration of FPGA into robots.

As a case study, the proposed component with hardwired image labeling on an FPGA is implemented on programmable SoC equipped with the ARM processor and FPGA logic. The ROS-compliant FPGA component on Xilinx Zynq-7020 performs 26 times faster than that of software with the ARM processor, and even 2.3 times faster than that of PC. Moreover, the total latency of the component was 1.7 times faster than that of processing with pure software. Therefore, the ROS-compliant FPGA component achieves remarkable performance improvement, maintaining high development productivity by cooperative processing of hardware and software.

Communication of ROS nodes occupies a lot of execution time in ROS-compliant FPGA component. From now on, another study is necessary for the reduction of ROS node's communication latency.

## Acknowledgment

This work was supported by SCOPE (0159-0112), SOUMU, Japan.

## References

[1] Robotics Society of Japan (ed.): "Robot technology", Ohmsha, Ltd, 2011.

[2] Open Source Robotics Foundation : http://wiki.ros.org/.

[3] OpenRTM-aist: http://www.openrtm.org/openrtm/ja.

[4] OROCOS: http://www.orocos.org/.

[5] Takuya Suzuki, Yusaku Yamazaki, Hakaru Tamukoh and Masatoshi Sekine : "AMobile Robot System using IntelligentCircuitin Silicon" IEICE Technical Report VLD2011−105 , CPSY2011−68 , RECONF2011−64, pp.83~88, 2012, in Japanese.

[6] Xillybus: http://xillybus.com/.

[7] Yujin Garage: http://garage.yujinrobot.com/

[8] iRobot: https://www.irobot-jp.com/product/comparison.html

[9] Nihon Binary: http://www.nihonbinary.co.jp/index.html

[10] Kazushi Yamashina, Takeshi Ohkawa, Kanemitsu Ootsu, Takashi Yokota : "Fundamental Design of ROS Compliant Components of Image Processing on an FPGA for Robotic Application", conference paper collection of The 77th National Convention of IPSJ , pp.1-169~1-170, 2015, in Japanese.

[11] Kazushi Yamashina, Takeshi Ohkawa, Kanemitsu Ootsu, Takashi Yokota : "ROS Compliant Componentizing of Image processing Hardware on a Programmable SoC", IEICE Technical Report, RESCONF2015-8, pp.41~46, 2015, in Japanese.

[12] Farzin Piltan, N. Sulaiman, M. H. Marhaban, Adel Nowzary and Mostafa Tohidian : "Design of FPGA-based Sliding Mode Controller for Robot Manipulator", International Journal of Robotics and Automation (IJRA), 2, 3, 173 – 194, 2011.

[13] Piltan, F., Rahmani, M., Esmaeili, M., Tayebi, M. A., Cheraghi, M. P. H., Rashidian, M. R., & Khajeh, A. : "Research on FPGA-Based Controller for Nonlinear System", International Journal of U-& E-Service, Science & Technology, Vol.8, No.3, pp.11-28, 2015.

[14] Sánchez-Solano, S., Cabrera, A. J., Baturone, I., Moreno-Velo, F. J., & Brox, M. : "FPGA implementation of embedded fuzzy controllers for robotic applications", Industrial Electronics, IEEE Transactions on, 54(4), 1937-1945.

[15] Tzuu-Hseng S. Li, Shih-Jie Chang and Yi-Xiang Chen : "Implementation of Human-Like Driving Skills by Autonomous Fuzzy Behavior Control on an FPGA-Based Car-Like Mobile Robot", IEEE Trans. on Industrial Electronics, 50, 5 , 867 – 880, 2003.

[16] Leong, Philip Heng Wai, and Kuen Hung Tsoi : "Field Programmable Gate Array technology for robotics applications.", Robotics and Biomimetics (ROBIO), IEEE International Conference on. IEEE, 2005.

[17] Takeshi Ohkawa, Daichi Uetake, Takashi Yokota, Kanemitsu Ootsu, Takanobu Baba : "Reconfigurable and Hardwired ORB Engine on FPGA by Java-to-HDL Synthesizer for Realtime Application,", Proc. 4th International Symposium on Highly Efficient Accelerators and Reconfigurable Technologies (HEART 2013), pp.45-50, 2013.